# FAST EMERGENCY PATHS SCHEMA TO OVERCOME TRANSIENT LINK FAILURES IN OSPF ROUTING

Fernando Barreto, Emílio C. G. Wille and Luiz Nacamura Jr.[1]

[1] Federal University of Technology - Paraná (UTFPR), Brazil

`fbarreto@utpr.edu.br, ewille@utfpr.edu.br, nacamura@utfpr.edu.br`

## ABSTRACT

*A reliable network infrastructure must be able to sustain traffic flows, even when a failure occurs and changes the network topology. During the occurrence of a failure, routing protocols, like OSPF, take from hundreds of milliseconds to various seconds in order to converge. During this convergence period, packets might traverse a longer path or even a loop. An even worse transient behaviour is that packets are dropped even though destinations are reachable. In this context, this paper describes a proactive fast rerouting approach, named Fast Emergency Paths Schema (FEP-S), to overcome problems originating from transient link failures in OSPF routing. Extensive experiments were done using several network topologies with different dimensionality degrees. Results show that the recovery paths, obtained by FEP-S, are shorter than those from other rerouting approaches and can improve the network reliability by reducing the packet loss rate during the routing protocols convergence caused by a failure.*

## 1. INTRODUCTION

The Internet is currently becoming a huge infrastructure to exchange any kind of IP traffic data such as email, web pages and P2P. This infrastructure has also been recently used for business applications, exchanging critical information such as electronic commerce, bank transactions and VoIP. This high dependence on the Internet is causing an increase of research for reliable IP networks. A reliable IP network must be able to sustain traffic flows, even when a failure occurs and changes the network topology. At the same time, the reliability of the Internet becomes crucial role for maintaining service performance.

Many applications executing over the Internet require a minimum quality of service in order to better fulfill customers' expectations. Examples of critical services include voice-over-IP (VoIP), electronic commerce, bank transactions, and video conferencing, among others. For those applications, networks need to guarantee very low packet loss rate, and low delays. At the same time, routing protocols should be able to achieve high network availability and a fast reaction time after a failure that changes the network topology occurs.

The majority of failures in real IP networks can be either of a single link, router or Shared Risk Link Group (SRLG) failure [1], [2], [3]. An SRLG is a group of links that share a common resource, such as a conduit or a line card. When a link fails, all the links belonging to the same SRLG should also be considered as failure. After a failure occurs, the main Interior Gateway Protocol (IGP), i.e., Open Shortest Path First (OSPF) [3], typically takes various seconds to identify the hardware failure, to signal this occurrence to all routers, to update their link state database, and finally to update their Forwarding Information Base (FIB). The FIB contains all the information necessary for a router to forward any packet. These reaction steps are fully described in Section 3 from [7] and Section 4.2.1 from [22]. This time-consuming reaction, named convergence period, may cause instability in the forwarding process with high packet loss rate, which consequently reduces network reliability. It is possible to reduce the convergence period to hundreds of milliseconds as in [5]. However, this can also result in routing instability [1], [6].

Recently, other approaches have focused on a proactive local recovery, defined as *IP Fast Rerouting* (IPFRR) [7]. These approaches help IP routing protocols (mainly the OSPF) reduce the routing instability caused by single failures. These mechanisms are proactive and identify in advance recovery paths that are intended to be temporarily used, since they are emergency solutions to be used against a component failure until the IGP converges.

There are various IPFRR approaches and a brief description of them is presented in Table 1. All these approaches can not avoid congestion caused by the deviated traffic over other traffics not originally affected by the failure, which can create more instability in the network. As only the NotVia [8] approach can reach 100% single-failure recovery (link, router or SRLG) even with asymmetric link weights, we explain it in details here.

Table 1. IPFRR approaches.

| Approach | Description | Drawbacks |
|---|---|---|
| FIR [2] [3] | Local rerouting with a forwarding process based on FIBs created per network interface. These FIBs are obtained by link failure deduction. | Works only for link failure. It can generate routing loops and does not support asymmetric links and SRLG. |
| FIFR [20] | Improves the [2] [3] approach in order to bypass router failure by deduction. | Cannot bypass a link failure adjacent to the destination router. Moreover, it still does not support asymmetric links and SRLG. |
| Loop Free Alternates (LFA) [11] | Defines a calculation to identify a backup neighbour router to bypass an adjacent failure in order to achieve alternative paths without routing loop. | Depends on the topology and does not guarantee 100% failure recovery. |
| FNH [21] | Identifies a *Feasible Next-Hop* (FNH) to bypass a link failure, and creates a *Rerouting Path* tunnel. | Only bypasses link failure and depends on a reactive process. |
| NotVia [8] | Finds routes to *not-via* addresses, which are used to encapsulate traffic flows compromised by a failure on a given network component. | Depends on the encapsulation technique, needs high extra information added to the FIB and can generate longer recovery paths. |
| Multiple Routing Configurations (MRC) [9] | Builds some sub-topologies, and each one can bypass some failure components. It selects one of them to bypass a failure. It needs one RIB and FIB for each sub-topology. | Adds high extra information to the FIB with various RIBs and FIBs. It can also generate unnecessary longer recovery paths and does not support SRLG. |
| U-Turn [12] | Locates a router ahead a neighbour router which is similar to the approach [11]. | Depends on the topology and does not guarantee 100% failure recovery. |
| Tunnels [10] | Builds tunnels reusing the existent routes from the source router until a router capable of bypassing an adjacent failure. | Does not support asymmetric links, does not guarantee 100% failure recovery and depends on the encapsulation technique. |

The NotVia approach defines *not-via* addresses, each one representing a component to be bypassed (failure). It also defines the *next-next-hops*, each one identifying the *not-via* address that will be used. If an adjacent component fails, the router searches for the *next-next-hop* address to bypass the failure. This search returns a *not-via* address that is used to encapsulate the packets in order to forward them to another router. Because all these *not-via* addresses are known by all routers, any router can identify which component has failed when it receives a *not-via* encapsulated packet, enabling a safe forwarding to bypass that failure until the packet reaches the router that owns this *not-via* address, where the packet is desencapsulated. However, this approach may need large FIB extensions to represent the *not-via* addresses and *next-next-hops* in the FIB. Moreover, it can generate longer recovery paths and the encapsulation process dependence can create packets that bypass the *Maximum Transfer Unit* (MTU) limit, which causes fragmentation in IPv4 networks (or even drop the packets if the *Don't Fragment* bit is set) and packet drop in IPv6. Another work reveals these same problems described for NotVia in [23].

This work proposes an IPFRR proactive based approach named Fast Emergency Paths Schema (FEP-S) in order to help OSPF during its convergence period. This approach generates shorter failure-recovery paths, each one identified as a Fast Emergency Path (FEP), which are added as small extensions in the FIB. The main contributions of this work, compared with the NotVia approach, are the small recovery path size of each FEP and the small extensions in order to represent the FEP generated in the FIB. During a failure, the FEP to be followed by the deviated packets are obtained from a simple packet mark whenever possible, and the use of encapsulation process is performed only in special cases (NotVia always requires the encapsulation process). Finally, FEP-S can reach 100% single failure recovery (link, router or SRLG) on a topology.

## 2. FAST EMERGENCY PATHS SCHEMA (FEP-S)

FEP-S is a distributed approach to generate failure recovery paths identified as Fast Emergency Paths (FEP). FEP-S is based on the *FEP-S Calculation* and on the *FIB Extension* (*FEP_Mark/NI* and *FEP_DifFor*). The *FEP-S Calculation* uses the OSPF metric (the shortest path from the source router to any destination router considering the sum of link weights) and, whenever possible, the number of routers in the path. The use of OSPF metric allows a reuse of the already installed OSPF routes, which reduces the recovery path calculation complexity. The number of routers provides a simple extra criterion to select a path among various paths with the same cost. All FEP used by a router are previously self generated and are represented directly at the FIB with an *FEP_Mark/NI* mark.

The *FEP_DifFor* process enables a failure-recovery process with the FEP represented at FIB by *FEP_Mark/NI*, as soon as an adjacent failure is detected.

To guarantee 100% single-failure recovery, there must be a physical topology still able to have all its routers connected in the presence of a single failure or SRLG failure. Besides, in order to enable a successful recovery, a network traffic distribution that occupies, at most, 50% link capacity is also necessary. This network traffic distribution is already being planned in real networks aiming to accommodate the deviated traffic during the presence of a failure [1]. These constraints can be adopted by any IPFRR approach; otherwise, neither is able to reach up to 100% single failure recovery, depending on the network topology structure.

### 2.1. FEP-S Calculation

Each router generates its own FEP and this calculation should follow the OSPF process execution aiming to reuse the link state database updated by the OSPF process.

Table 2 presents the notations used in this calculation. Consider a network topology represented by a graph G(V,A), where V is the set of routers and A is the set of links that connect the routers. Consider $(i,j) \in A$ the representation of a link connecting routers $i$ and $j$. For each link $(i,j)$, a cost $c_{i,j}$ is specified during the OSPF configuration.

Table 2. Notation used by FEP-S Calculation.

| Symbol | Description |
|---|---|
| $SR$ | Source Router |
| $AR$ | Adjacent Router to $SR$ to be bypassed (supposedly with failure) |
| $NR$ | Neighbour Router to $SR$: first router of $ALT_{SR,DR,AR}$ |
| $DR$ | Destination Router |
| $(SR,AR)$ | Adjacent Link to $SR$ to be bypassed (supposedly with failure) |
| $Z_{SR,J,AR}$ | Cost function of the shortest paths from $SR$ to any other destination router $J$ |
| $\varphi_{SR,J,AR}$ | Set of shortest paths to each $J$ router which obey to $Z_{SR,J,AR}$, and exists some shortest paths which are alternative paths to some $J$ routers |
| $\varphi_{SR,DR,AR}$ | Set of shortest alternative paths to $DR$, from $\varphi_{SR,J,AR}$ |
| $ALT_{SR,DR,AR}$ | A shortest alternative path from $\varphi_{SR,DR,AR}$ |
| $RF$ | *Router FEP*: router belonging to $ALT_{SR,DR,AR}$ which identifies the last router in $FEP_{SR,DR,AR}$ |
| $FEP_{SR,DR,AR}$ | Sub-path (FEP) from $SR$ until $RF$ from $ALT_{SR,DR,AR}$, candidate to be $S\_FEP_{SR,DR,AR}$ |
| $S\_FEP_{SR,DR,AR}$ | FEP chose for $DR$ which obey to $Z'_{SR,DR,AR}$ |
| $Z'_{SR,DR,AR}$ | Cost function which minimizes the cost and the number of routers in each $FEP_{SR,DR,AR}$ |
| $num\_FEP_{SR,DR,AR}$ | Represents the number of routers agreeing to the specified level (ECMP, LFA or SIG) |
| $cost\_FEP_{SR,DR,AR}$ | Represents the sum of link costs agreeing to the specified level (ECMP, LFA or SIG) |
| $J$ | Set of destination routers from $SR$ |
| $N$ | Number of routers in a topology |
| $c_{i,j}$ | Link cost of link $(i,j)$ |
| $x_{i,j}$ | Link flow at link $(i,j)$ |
| $K$ | Each router analyzed to be $RF$ from $ALT_{SR,DR,AR}$ router sequence |
| $DistOSPF(x,y)$ | Distance of OSPF shortest path (sum of link costs) from router $x$ to $y$ |
| $NumRoutersOSPF(x,y)$ | Number of routers in the OSPF shortest path from router $x$ to $y$ |
| $DistFEP_{SR,DR,AR}$ | Sum of link costs from $FEP_{SR,DR,AR}$ |
| $NumRoutersFEP_{SR,DR,AR}$ | Number of routers in $FEP_{SR,DR,AR}$ |
| $OSPF\_Links(x,y)$ | Set of links belonging to the OSPF shortest path from $x$ to $y$ |
| $OSPF\_Routers(x,y)$ | Set of routers belonging to the OSPF shortest path from $x$ to $y$ |

The SPF calculation executed by a *Source Router* (SR) minimizes the sum of link costs $c_{i,j}$ from $SR$ to any other router in the topology. The first FEP-S mathematical formulation is based on linear programming for the *Shortest Path Problem* [13]. The FEP-S calculation finds the cost $Z_{SR,J,AR}$, which minimizes the sum of link costs respecting the number of unit flows that passes through the links in order to discover the shortest paths from $SR$ to a set of destination routers $J$. The FEP-S formulation adapts [13] to avoid adjacent components to $SR$ supposedly with failure: *Adjacent Router* (AR) or *Adjacent Link* (SR,AR) or SRLG.

$$Z_{SR,J,AR} = min \sum_{i=1}^{n} \sum_{j=1}^{n} c_{i,j} x_{i,j}$$

Subject to:

$$\sum_{j=1}^{n} x_{i,j} - \sum_{k=1}^{n} x_{k,i} = \begin{cases} n\text{-}1, & if \ i = 1 \\ \text{-}1, & if \ i \neq 1 \end{cases} \tag{1}$$

$$c_{SR,j} = \infty, \ if \ j{=}AR \ and \ if \ J{=}AR \tag{2}$$

$$c_{i,j} = \infty, \ if \ j{=}AR \ and \ if \ J{\neq}AR \tag{3}$$

$$c_{i,j} = \infty, \ if \ (i,j) \in \begin{cases} SRLG(SR,AR) \, , \ if \ J \ = \ AR \\ SRLG(AR, ?) \, , \ if \ J \ \neq \ AR \end{cases} \tag{4}$$

$$i \geq 0, \ \ j{\geq} 0, \ i \neq j, \ x_{i,j} {\geq} 0, \ c_{i,j} {\geq} 0 \tag{5}$$

The constraint (1) selects the shortest path transmitting *n-1* units of flow from *SR*, which means, for every router reached in the shortest path, 1 unit is consumed. Constraints (2) and (3) avoid shortest paths, which use (*SR,AR*), if *J=AR*, or use *AR* (all the adjacent links to *AR*), if *J≠AR*, since they set the link cost with an ∞ value. The formulation always tries to bypass *AR* (for all *J≠AR*), because during the indication of a real hardware failure, *SR* cannot immediately identify whether the failure is from (*SR,AR*) or *AR*. Thus by choosing *AR*, FEP-S will consequently bypass (*SR,AR*). The links belonging to the set *SRLG(SR,AR)*, if *J=AR*, or the links belonging to the set *SRLG(AR, ?)*, where "*?*" means any router adjacent to *AR* if *J≠AR*, are considered in (4), which sets the link cost with an ∞ value to avoid their use in the shortest path. The solution for this problem can be obtained from a simple modification on Dijkstra´s algorithm to consider constraints (2), (3) e (4). In addition, considering that only a segment of the SPF tree is affected, the use of *Incremental-SPF* [14] approach can reduce this time-consuming process considerably [5].

The shortest paths that obey $Z_{SR,J,AR}$ are represented in the set $\varphi_{SR,J,AR}$ . Consider the shortest paths to a Destination Router (*DR*) belonging to $\varphi_{SR,J,AR}$, where *DR* belongs to *J*. If the original OSPF path from *SR* to *DR* (considering all topology components) uses at least (*SR,AR*), then, the obtained shortest paths with $Z_{SR,J,AR}$ are alternative paths to *DR* and are candidates to bypass (*SR,AR*) or *AR*, which will depend on the *DR* location.

The set of possible alternative paths to *DR* with these characteristics are represented as $\varphi_{SR,DR,AR}$. Each alternative path from $\varphi_{SR,DR,AR}$ is represented as $ALT_{SR,DR,AR}$.

The second FEP-S mathematical formulation below ($Z'_{SR,DR,AR}$) was created to identify, in each $ALT_{SR,DR,AR}$ , which router is able to bypass the failure component ((*SR,AR*) or *AR*) when, from this router, the original OSPF path is not affected by the failure component and can safely reach *DR*. This identification process adapts the proposal from [7] in order to use classification levels (*Equal Cost MultiPath* (ECMP), *Loop Free Alternates* (LFA) or *Signaling* (SIG)) and facilitate the discovery of this router, which is named *Router FEP* (*RF*). Each router analyzed to be *RF* from $ALT_{SR,DR,AR}$ router sequence is named *k*. The ECMP level finds the *RF* at the first router (*Neighbor Router - NR*) of $ALT_{SR,DR,AR}$ , and $ALT_{SR,DR,AR}$ must have the same OSPF distance cost (*DistOSPF*) as the original OSPF path from *SR* to *DR*. The LFA level adapts the original LFA approach [11] to also find the *RF* at *NR*, and adds the number of routers constraint to select $ALT_{SR,DR,AR}$ . The SIG level identifies the *RF* always after *NR*, so *k* goes through the $ALT_{SR,DR,AR}$ router sequence until a router with the *RF* characteristics is located. All *k* which cannot be *RF* are stored in the set named *intermediary_routers*. Thus independently of the level used, a sub-path is generated for each $ALT_{SR,DR,AR}$, from *SR* to *RF*, named $FEP_{SR,DR,AR}$. All these $FEP_{SR,DR,AR}$ for $\varphi_{SR,DR,AR}$ are candidates to be the selected FEP for *DR*, which is denoted as $S\_FEP_{SR,DR,AR}$.

In order to obtain $S\_FEP_{SR,DR,AR}$ , we find the cost $Z'_{SR,DR,AR}$ which minimizes the sum of *1000 cost_FEP$_{SR,DR,AR}$* and the number of routers *num_FEP$_{SR,DR,AR}$*.

$Z'_{SR,DR,AR} = min(1000cost\_FEP_{SR,DR,AR} + num\_FEP_{SR,DR,AR})$

Subject to:

$\varphi_{SR,DR,AR} \subset \varphi_{SR,J,AR}$ (6)

$ALT_{SR,DR,AR} \in \varphi_{SR,DR,AR}$ (7)

$FEP_{SR,DR,AR} \subset ALT_{SR,DR,AR}$ (8)

$RF=k$, if $DistOSPF(SR,k) + DistOSPF(k,DR) = DistOSPF(SR,AR) + DistOSPF(AR,DR)$ (9)
    and $k \in Neighbors\_SR$

$RF=k$, if $DistOSPF(k,DR) < DistOSPF(k,SR) + DistOSPF(SR,DR)$ (10)
    and $k \in Neighbors\_SR$

$RF=k$, if $DistOSPF(k,DR) < DistOSPF(k,SR) + DistOSPF(SR,DR)$ (11)
    and $k \notin Neighbors\_SR$

$AR \notin OSPF\_Routers(k,DR)$, if $AR \neq DR$ (12)

$OSPF\_Links(k,DR) \cap \begin{cases} SRLG(SR,AR)=\varnothing, \text{ if } AR=DR \\ SRLG(AR,?\,)=\varnothing, \text{ if } AR \neq ND \end{cases}$ (13)

$k \in FEP_{SR,DR,AR}$ , if no constraint in sequence (9), (10) or (11) is possible (14)

$k \in FEP_{SR,DR,AR}$ and it is the last router, if $(RF==k)$ & (a constraint in sequence (9), (10) (15)
    or (11) is possible)

$num\_FEP_{SR,DR,AR} = \begin{cases} 1, \text{ if constraint (9) and (15) is possible} \\ NumRoutersOSPF(k,DR), \text{ if constraint (10) and (15) is possible} \\ NumRoutersFEP_{SR,DR,AR}, \text{ if constraint (11) and (15) is possible} \end{cases}$ (16)

$cost\_FEP_{SR,DR,AR} = \begin{cases} DistOSPF(k,DR), \text{ if constraint ((9) or (10)) and (15) are possible} \\ DistFEP_{SR,DR,AR}, \text{ if constraint (11) and (15) are possible} \end{cases}$ (17)

$k \in ALT_{SR,DR,AR}$ ; $AR \notin Neighbours\_SR$ (18)

The number of routers only influences the minimization process if there is more than one FEP with the same $cost\_FEP_{SR,DR,AR}$, due to the *1000* multiplying factor. The localization of *RF* is obtained via analysis of each router of $ALT_{SR,DR,AR}$ router sequence, being each router in its turn of analysis represented by $k$. Variable $k$ turns to be *RF* and it is the last router in $FEP_{SR,DR,AR}$, i.e. constraint (15), and only if one of the constraints, in this sequence, is accepted: (9) for ECMP level, (10) for LFA level or (11) for SIG level. In SIG level, because $k$ is a router after *NR*, it does not belong to the *SR* neighbor routers set (*Neighbors_SR*), i.e. constraint (11). In case none of these three constraints is accepted, $k$ belongs to $FEP_{SR,DR,AR}$, i.e. constraint (14), then another $k$ from $ALT_{SR,DR,AR}$ router sequence must be analyzed. Constraint (12) avoids $ALT_{SR,DR,AR}$ if an OSPF router sequence path exists from $k$ to *DR* which uses *AR*, if $AR \neq DR$. Constraint (13) avoids $ALT_{SR,DR,AR}$ if an OSPF link sequence path exists from $k$ to *DR* which uses some link belonging to an SRLG. Constraint (10) defines a value for $num\_FEP_{SR,DR,AR}$ according to the level chosen: 1 for ECMP (9), number of routers in the OSPF path, i.e. *NumRoutersOSPF*, for LFA (10) or number of routers in the FEP path (from *SR* to *RF*), i.e. *NumRouterFEP*, for SIG (11). If the ECMP level is chosen, it sets the value 1 in order to allow multiple paths and load balancing with ECMP. Constraint (17) also defines a value for $cost\_FEP_{SR,DR,AR}$ according to the chosen level: OSPF distance from $k$ to *DR* for ECMP and LFA levels, and FEP distance (*DistFEP$_{SR,DR,AR}$*) from *NR* to *RF* for SIG level.

The formulation enables each router in a topology to select the $S\_FEP_{SR,DR,AR}$ from all possible alternative paths from $\varphi_{SR,DR,AR}$ to each *DR*. The proactive calculation of FEP based on *RF* location previously provides an $S\_FEP_{SR,DR,AR}$ able to deviate the packets to a recovery path towards *DR* when an adjacent failure occurs. If a failure is at the *AR*, this recovery path is always the same one that would be used by the packets, after the OSPF reaction, from the router closest to this failure. But if a failure is at the (*SR,AR*), then the recovery paths are not always the same. This occurs because the FEP proactive calculation always tries to bypass *AR*, if $AR \neq DR$, in order to achieve more effective recovery path (a router can not immediately identify if an adjacent failure signal is from (*SR,AR*) or from *AR*). The $S\_FEP_{SR,DR,AR}$ selected for each *DR* are stored in FEP_Vector[*DR*]. In this sense, this calculation improves the previous FEP-S calculation presented in our work [24] by completing the mathematical formulation.

## 2.2. FEP-S FIB Extension

Once the $S\_FEP_{SR,DR,AR}$ to all *DR*s have been generated by each router and stored in FEP_Vectors, the FEP-S must then add an extension at the FIB to represent which FEP to be used when a real failure occurs. All related approaches also have a similar extension, but the FEP-S extension consists of a pair packet mark (*FEP_Mark*), Network Interface (*NI*), and of a differentiated forwarding (*FEP_DifFor* process).

### 2.2.1. *FEP_Mark/NI* Mark

A *FEP_Mark* is a packet mark defined by each *SR* to identify each FEP generated. It is represented with 16 bits divided into nine bits to identify the *SR* (*SR id*), and seven bits to identify each FEP (*FEP id*). The *SR id* achieves up to 512 IGP routers, with at most 128 different *FEP id* to be used per router. The limit of 512 routers is more than sufficient to represent a real topology using IGP. In practice, a topology with more than 512 routers in a single area could not be considered a scalable topology to link state routing protocols due to the high control traffic they need to exchange. For the *FEP id*, in our tests with various real topology representations, we reach at most 42 different FEP and an average 19. Then, we believe, seven bits are sufficient to represent all FEP id per *SR*. If necessary, the number of bits for *SR id* and *FEP id* in 16 bits can be managed in order to achieve more or less *FEP id* than *SR id*, but this change must be planned and defined in all IGP routers with the same pattern.

In order to represent the 16 bits *FEP_Mark*, both at the IPv4 and at the IPv6 headers, FEP-S reuses some fields from each IP header version:

- IPv4: The Fragment Offset field can be reused if the Path MTU Discovery [15] is used.

- IPv6: The Flow Label field can be used when setting the first three bits with "111". It identifies the remaining 17 bits for future use [16], though we reuse 16 bits.

An *NI* is represented with 8 bits, which can represent up to 255 network interfaces per router. This number is high enough for today's common hardware routers. If necessary, the number of *NI* bits can be increased to identify more network interfaces per router, but it increases the amount of extra information added to the FIB.

One *FEP_Mark/NI* pair can be generated for each FEP_Vector[*DR*] obtained, but in order to reduce the number of *FEP_Mark/NI* generated it must initially follows two rules:

- All existing FIB entries with network prefixes announced by the same *DR* (information obtained from the link state database) use the same FEP_Vector[*DR*]. In this case, these FIB entries must refer to one same *FEP_Mark/NI* pair, because the packets will follow the same FEP_Vector[*DR*] to *DR*.

- In case there are two or more FEP_Vectors with different *DR*s, however with the same routers sequence, then all FIB entries with network prefixes announced by these *DR*s must refer to the same *FEP_Mark/NI* pair. This is possible because the FEP will be the same for the different *DR*s and they need one mark only.

After observing these two rules, the *FEP_Mark/NI* pairs are then generated and added to the FIB following one of the two following cases: *SR case* or *Not_SR case*.

The *SR case* occurs when a router generates its own FEP_Vectors. It is thus the *SR* in the *FEP-S Calculation*. In this case, for each FEP_Vector[*DR*], an *FEP_Mark* is generated with an "*SR id*" corresponding to the last 9 bits of the *SR* IP loopback address (*Router ID*) [4], which is possible, because they are usually organized within the same IP group address projected to be used by the IGP routers. If this simple IP loopback address organization is not adopted, the FEP-S approach can not work properly. If the FEP_Vector[*DR*] is obtained via ECMP or LFA level, then the *FEP id* has seven bits with "0" value. Otherwise (SIG level), the *FEP id* is generated during the SIG level calculation (an incremental identified number for each SIG FEP_Vector[*DR*]). The *NI* to be selected is the one connected to the first router of a FEP_Vector[*DR*], i.e., *NR*.

The *Not_SR case* occurs when a router belongs to a FEP_Vector[*DR*] originated by any other router, say *o-r*, through SIG level, and in this case, the router can be the *NR, intermediary_routers* and *RF* from a FEP_Vector[*DR*], generated at *o-r*. In *Not_SR case*, a router generates the *FEP_Mark/NI* pair using mainly the FEP_Vector[*DR*] and the *FEP_Mark* information obtained from *o-r* through *FEP_Signal* process, which is detailed in the next section. With this information available, each router in *Not_SR case* (*NR, intermediary_routers* and *RF*) only needs to identify the *NI* to complete the pair *FEP_Mark/NI*. To conduct this process, consider a router Y, as one of *Not_SR case* routers, which can simply identify the *NI* as the one connected to the next router after Y in the FEP_Vector[*DR*] routers sequence.

Figure 1 illustrates how these pairs are represented in the FIB with reduced number of *FEP_Mark/NI* generated. There are three pairs of *FEP_Mark/NI*. The two first *FEP_Mark/NI* entries are generated by the *SR case*. We show two of the FIB entries with destination addresses (X and Y) announced by the same router (D). These two entries refer to the same *FEP_Mark/NI* pair (*FEP_Mark1/NI*), which reduces the resource use. There are also two FIB entries with destination addresses (T and Z) announced by different routers (M and N). These two entries refer to a same *FEP_Mark/NI* (*FEP_Mark2/NI*). The reference from a FIB entry to the *FEP_Mark/NI* is achieved with an additional field named *Ref* of 8 bits, which is expected to be sufficient to refer to all *FEP_Mark/NI* pairs generated by the *SR case*. In our experiments, involving various topology representations, a router needed at most 6 bits for *Ref*, though an average of all routers shows that 4 bits were sufficient. The *FEP_Mark/NI* pairs generated by *Not_SR case* are only added to the FIB but not referenced by any FIB entry, as show by the *FEP_Mark3/NI* pair in Figure 1. These *Not_SR case* marks are only used when an already marked packet (marked by a *SR*) arrives at the router.

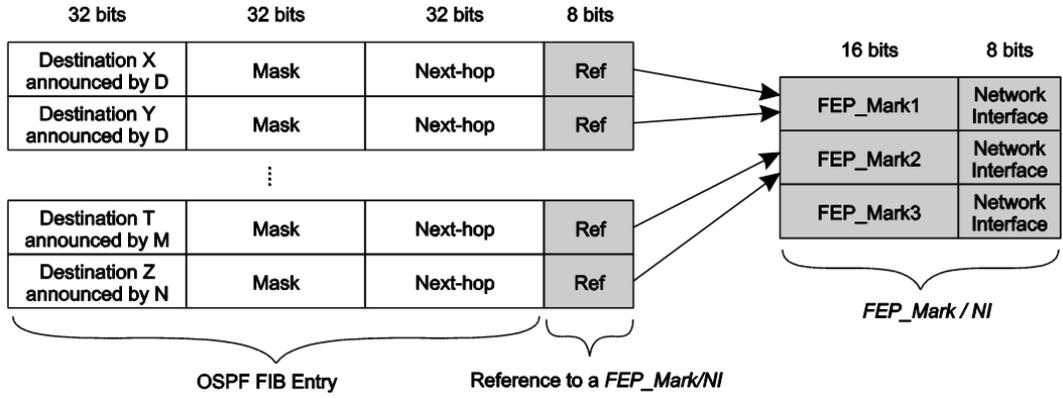

Figure 1. *FEP_Mark/NI* Extension at the FIB

Only the OSPF network prefixes must be updated with these pairs, because the BGP network prefixes at FIB (if exist) use the IGP routes to provide BGP next hop reachability. In case of multiple OSPF areas, the FEP-S uses the link state database of an area, and each *Area Border Routers* announces its summarized external network prefixes in the area. These network prefixes are handled in the same way as the prefixes announced by any DR.

With these *FEP_Mark/NI* pairs added, the FEP_Vectors can be used during the occurrence of a failure and this is accomplished by the *FEP_DifFor* process.

### 2.2.2. *FEP_DifFor* Process

Only the SR is capable of marking the packets (IPv4 or IPv6 headers) with *FEP_Mark* to indicate which specific FEP_Vector[*DR*] must be used to bypass an adjacent failure. This process is performed by a differentiated forwarding defined as *FEP_DifFor* which is appended to the original OSPF forwarding process. If there is no adjacent failure, the original OSPF forwarding process is always used. If there is an adjacent failure ((*SR,AR*) or *AR*), the router acts as a *SR* and the *FEP_DifFor* tries to recover the traffic flows affected by the failure using the *FEP_Mark* referenced at the FIB in order to mark the packets and forward them to the respective *NI*. In IPv4 networks, *FEP_DifFor* also sets the first bit of Flags field to "1". This is necessary to identify the content of the *Fragmentation Offset* field as generated by FEP-S approach (not from IPv4 fragmentation process) and to inform that the packet is being deviated from a failure. This bit can be changeable without a problem, in this case, because the *FEP_DifFor* assumes the IGP forwarding process from the time the *FEP_Mark* is found in the packet. In IPv6 networks, *FEP_DifFor* also sets the first "111" bits which can already inform other routers that the packet is being deviated, since it is not used by IPv6 *Flow Label* approach [16].

In addition, the *FEP_DifFor* performs the forwarding in each router, from *SR* to *RF*, not only based on the IP destination address, but also on the *FEP_Mark*. The *FEP_Mark* identifies which NI has to be used from the respective *FEP_Mark/NI* pair. This process allows the deviated packets to follow the correct FEP_Vector routers sequence. Thus, from *RF*, the normal OSPF forwarding is thereafter used by *FEP_DifFor* to forward the packets based on the destination IP until they reach the *DR*. At *DR*, the added mark is removed and the first bit of Flags (IPv4 networks) is restored to its previously standard value: "0". It is important to notice that, during the deviation, the *FEP_DifFor* does not perform a *FEP_Mark* switch and it does not act when the packets are unmarked (original OSPF forwarding is used, instead). As other related approaches, this schema requires that all routers in a recovery path are configured with FEP-S in order to handle the marked packets.

During a recovery process, if the 16 bits are already being used (IPv4 normal fragmentation process with *Don´t Fragment* bit with "0" or IPv6 *Flow Label* without "111" first bits set), the *FEP_DifFor* can bypass this problem through the encapsulation process. *FEP_DifFor* will encapsulate the packets with the *DR* address to add the *FEP_Mark*. The packets will follow the same FEP_Vector as the original IP address destination, since the original destination network address is announced by this same *DR*. At *DR*, they are desencapsulated. This method ought to be used only when the 16 bits are previously set. Besides, we believe that these 16 bits have not been used today by normal network traffic, because the *Fragment Offset* field will not be used due to the *Path MTU Discovery* process, which is widely disseminated in IPV4 networks (only 0,06% of backbone traffic has fragmentation [19] ). Also, in IPv6 networks, the typical *DiffServ* mechanisms use the *Class of Service* field instead of *Flow Label*.

In the presence of multiple independent failures (not SRLG), the NotVia [8] approach is currently unable to identify this occurrence, which tends to generate network instability. The original LFA approach [11], in this case, reveals the possibility of routing loop. Otherwise, the *FEP_DifFor* can identify this occurrence through the mark added to the deviated packets, which remains marked until they reach *DR* (if multiple failures occur, these marked packets are dropped to avoid a new FEP action which will cause instability during the OSPF convergence). From *RF*, the *FEP_DifFor* routes the packets based only on the normal OSPF forwarding until the packets reach *DR*. If any new failure occurs through the recovery path until *DR*, the marked packets are dropped to avoid another recovery process. Thus, when a router receives a marked packet (*FEP_Mark*), and the *NI* (using *FEP_DifFor*) or the *next-hop* (*FEP_DifFor* routing based on the normal OSPF forward) is unavailable, the router drops the packet.

### 2.2.3. *FEP_Signal* Process

For the FEP_Vectors generated by ECMP or LFA, the *FEP_DifFor* needs to deviate the packets based on *FEP_Mark* until *NR*. However, the FEP_Vectors generated by SIG level extend to a respective *RF* beyond *NR*. The *FEP_DifFor* needs to forward the marked packets to correctly reach the *RF* during a failure, and avoid possible unstable routing caused by pure OSPF forwarding (routers which have not been converged yet). The information required by the *FEP_DifFor* to perform this process is obtained from *FEP_Signal*.

The *FEP_Signal* is a proactive simplified signaling extension to OSPF. Each *SR* uses the selected FEP_Vectors (described in previous section) to identify the FEP routers (*NR*, *intermediary_routers* and *RF*) that should be informed with an *FEP_Signal* message. This message uses the well-defined generic *Link-State Advertisements* (LSA) already available at OSPF (*Opaque LSA* [15]), which enables support for future extensibility of OSPF. A *Link-State Type 9 Opaque LSA* packet is generated at *SR* containing the FEP_Vector, *SR*, *DR* and the generated *FEP_Mark*. This packet is sent through a link-local scope (type 9) to the network interface connected to the first router in FEP_Vector router sequence, i.e. *NR*. The *Opaque LSA* type 9 provides a minimal traffic necessary to inform each router from the FEP_Vector. When this packet is received at *NR*, it is analyzed by FEP-S process to configure the *FEP_Mark/NI* (using *Not_SR case* previously described) for this FEP_Vector (obtained from the *Opaque LSA* packet). In the sequence, FEP-S updates the packet as a new type 9 *Opaque LSA* and forwards it through a link-local scope to the network interface connected to the next router in the FEP_Vector sequence. This forwarding process occurs until this packet reaches *RF* (the last router pointed in FEP_Vector). At *RF*, the packet is sent back through the reverse FEP_Vector to be acknowledged by the FEP routers until it reaches *SR*. The return of this packet represents an acknowledgement to the *FEP_Signal* that this FEP_Vector has been successfully defined. Then, the *FEP_Signal* can discard this packet in the sequence. This process is repeated with all the selected FEP_Vectors. Thus, this proactive process is important to previously publish the *FEP_Mark* generated by *SR* to *NR*, *intermediary_routers* and *RF*, and to configure these routers to identify which *NI* should be used for a packet with *FEP_Mark*. The *FEP_Signal* can replace the *FEP-SE* process presented in our previous work [25] in order to reduce extra multiple calculation process to identify remote FEP_Vectors.

### 2.2.4. Maintenance of *FEP_DifFor* deviation

When a failure is detected, the *FEP_DifFor* is activated at *SR* to mark the packets with *FEP_Mark* and forward to the respective *NI*, apart from the OSPF next-hop. The *FEP_DifFor* remains with this deviating process for a period of time long enough for all FEP_Vector routers finish the execution of their OSPF convergence. This interval can be more precisely set according to the interval generated by the *oFIB* approach [17]. The *oFIB* stipulates a sufficient interval to execute the OSPF reaction at the router closest to a failure, i.e. *SR*, always after the routers further away from this failure have completed their OSPF reaction. At the end of this interval, the router closest to the failure (*SR*) is able to execute its OSPF reaction, and once this process is complete, the *FEP_DifFor* deviation process is disabled because the router closest to the failure use the correct updated routes with OSPF forwarding. In sequence, a new *FEP-S Calculation* is executed in background to obtain new FEP_Vectors according to the updated link state database.

During the deviation process, in case the packets represent a threat to other traffic flows not affected by the failure, the *FEP_DifFor* can attempt to avoid further complications by adopting the simple rule: only continue forwarding the deviated marked packets if the queue length is shorter than 80%. *FEP_DifFor* will drop the deviated packets for any value above this. In all the simulated tests, the 80% queue length value was proved to be enough to avoid harming the normal traffic flows when FEP is being used. This option is projected to be enabled/disabled by the network administrator for some selected traffic flows.

## 3. PROPOSAL EVALUATION

The FEP-S has been developed and implemented in Java Simulator (J-SIM - *http://sites.google.com/site/jsimofficial/*). We chose the J-SIM because its OSPF source code is adapted from Zebra GNU project (*http://www.zebra.org*). The Zebra GNU project provides free routing software, which has an OSPF protocol behaviour that conforms to the OSPF protocol [4]. The complexity of the FEP-S calculation is O($n^3.log(n)$), where *n* is the number of routers at the network topology. A description of this algorithm is provided in the Appendix. However, this complexity is limited in practice because, in an OSPF area, the number of routers can reach, at most, up to a few hundreds. Moreover, the algorithm reuses the OSPF link state database with *Incremental* SPF [14] and it is always executed in background only at very few times (only after an OSPF execution caused by a change in the link state database, e.g, a new router/link added or a link cost change), which should not compromise the router performance. As other IPFRR approaches, the FEP-S needs some modifications on the forwarding process to add the *FEP_DifFor* process. The *FEP_DifFor* needs only a few changes of verification in the normal OSPF forwarding to consider the *FEP_Mark/NI*, as explained in Section 2.2.

In addition, only the calculation of NotVia recovery paths was also implemented in J-SIM following the semantic described in [8], since this has been the only one to attain 100% single failure coverage (link, router or SRLG).

In order to evaluate the FEP-S approach compared with NotVia, nine well-known real backbone network topologies were used (Aleron, AGIS, Arpanet, BTN, CAIS, Electric LightWave (ELW), GEANT2, Level 3 and Sprint). The first evaluation of the recovery paths is based on the evaluation conducted in [18], which confronts the recovery paths extension in terms of the number of routers used among all source-destination pairs. This comparison allows a measure of the extension followed by the deviated packets. The NotVia approach does not have a native support from ECMP and LFA adapted to obtain their paths, as it is the case of the FEP-S. NotVia simply indicates the approach from [11] to be used in these cases [8]. However, in case of various LFA paths possible, the original LFA approach does not identify which one should be used and it only recommends the choice for anyone. Conversely, the FEP-S can improve the identification of the path with the smallest number of routers in case there are various LFA paths with same cost. This enables a reduction of the deviation path. Figure 2 shows the results of this first evaluation for the topologies. In all topologies, FEP-S frequently achieves shorter recovery paths than NotVia because the packets deviated with NotVia must follow a path to *next-next-hop*, and only from this router, they follow the correct path to destination. This has higher occurrence at Aleron, AGIS, BTN, CAIS and Level 3 topologies, because they have longer sequences of routers in serial, which forces the NotVia to deviate the packets to a longer unnecessary path in order to reach *next-next-hop* and then return. Conversely, this problem does not occur in FEP-S, because the packets are deviated via *FEP_Mark/NI* until *RF* only, which is in the shortest alternative path. The FEP-S generates almost the same router sequence than NotVia when the topology has a higher connectivity among the routers. This occurrence can frequently result in calculus for NotVia and FEP-S approaches where the *next-next-hop* and the *RF* respectively are in the same router. The Arpanet, Electric LightWave, GEANT2 and particularly at the Sprint backbone have this characteristic. The Sprint backbone presents the highest connectivity among the tested topologies, which justifies the almost equal recovery paths extension generated by both approaches.

The packet deviation generated by the *FEP_DifFor* itself follows the FEP_Vectors routers until RF. The extension in number of routers from the FEP_Vectors and the *FEP_Mark/NI* generation are minimized in the FEP-S formulations, which influence the amount of information added to the FIB. The second evaluation is the amount of information added to the FIB by FEP-S and NotVia approaches. A description of extra information added by these approaches and their representation is presented in Table 3.

Table 3. Extra FIB information for each approach (FEP-S and NotVia).

|  | FEP-S | NotVia |
|---|---|---|
| Extra Information to be added in FIB | *FEP_Mark/NI* (FNI) obtained according to Section 2.2 and referenced (*Ref*) in *OSPF FIB Entries* (OFE). The OFE are network prefixes only announced by routers in the OSPF area domain. | Each router adds a new FIB entry for each *not-via* address published by all the other routers in a OSPF area domain. It also adds one *next-next-hop* for each FIB entry. |
| Amount of bytes to represent the information | - *Ref* = 1 byte (Section 2.2) <br> - *FEP_Mark/NI* (FNI) = 3 bytes (Section 2.2) | - Each new *not-via* address entry in FIB (nFIB) uses Destination IP (4 bytes) + Netmask (4 bytes) + next-hop (4 bytes) = 12 bytes [4] <br> - Each *next-next-hop* (n-n-h) is one *Router ID* [8] = 4 bytes |

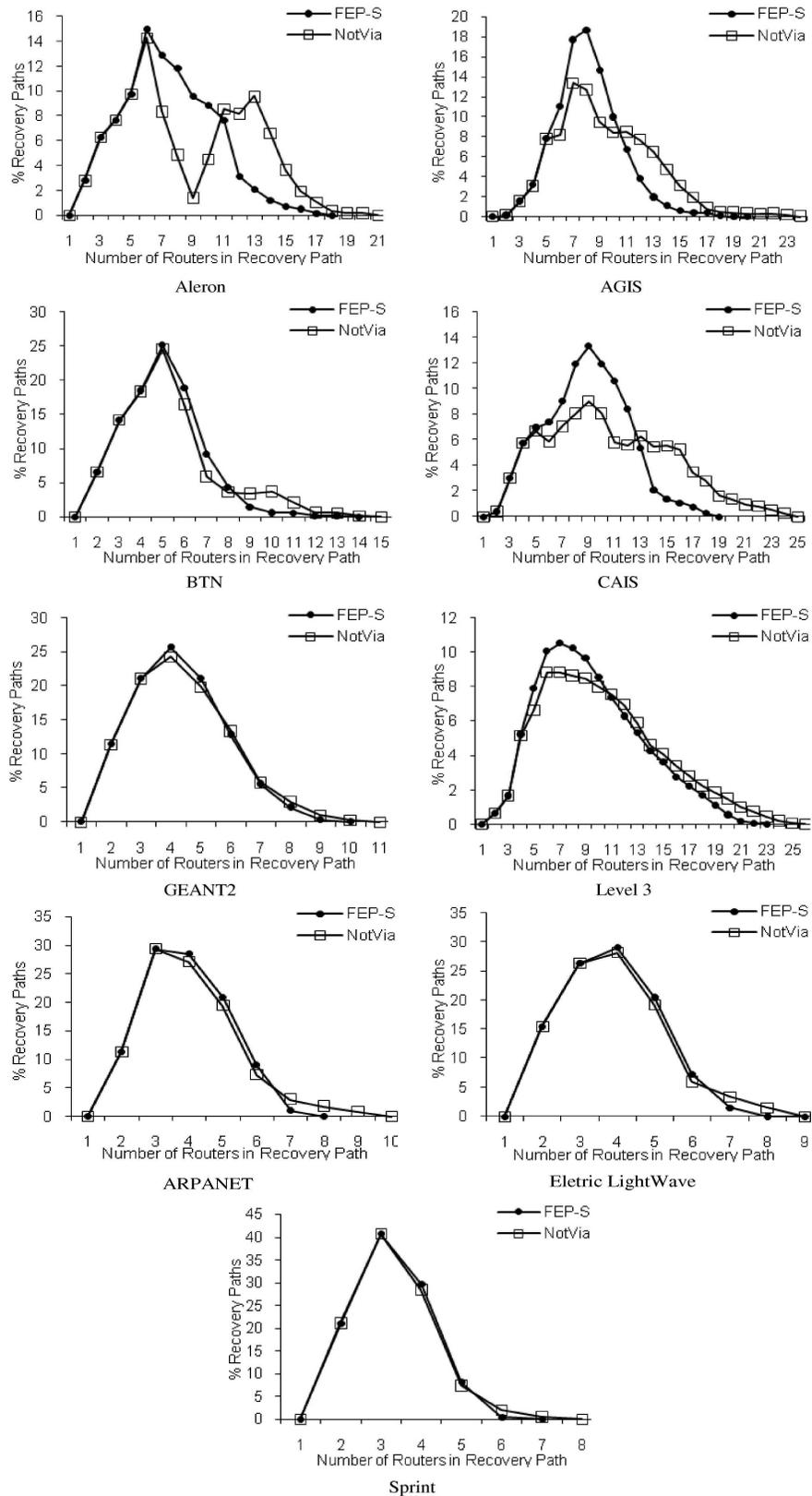

Figure 2. Number of routers in recovery paths per network topology

Table 4 shows a stipulated amount of extra information added to the FIB per router based on the Table 3 information. The FEP-S minimization process results in reduced information compared with NotVia, because it only needs the *FEP_Mark/NI* referenced among the *OSPF FIB Entries* with the *Ref* field, while the NotVia needs all *not-via addresses* added as new FIB entries and the *next-next-hop* added per FIB entry [8].

Table 4.  Extra FIB information added per router.

| Topology | FEP-S | NotVia |
|----------|-------|--------|
| Aleron | 27 FNI + OFE (*Ref*), *Total*: 81+ OFE bytes | 54 nFIB + OFE (n-n-h) *Total*: 648 +4(OFE) bytes |
| AGIS | 31 FNI + OFE (*Ref*) *Total*: 93 + OFE bytes | 149 nFIB +OFE (n-n-h) *Total*: 1788 +4(OFE)bytes |
| Arpanet | 9 FNI + OFE (*Ref*) *Total*: 27 + OFE bytes | 64 nFIB + OFE (n-n-h) *Total*: 768 +4(OFE) bytes |
| BTN | 15 FNI + OFE (*Ref*) *Total*: 45 + OFE bytes | 80 nFIB +OFE (n-n-h) *Total*: 960 +4(OFE) bytes |
| CAIS | 30 FNI + OFE (*Ref*) *Total*: 90 + OFE bytes | 88 nFIB +OFE (n-n-h) *Total*:1056 +4(OFE) bytes |
| ELW | 11 FNI + OFE (*Ref*) *Total*: 33 + OFE bytes | 62 nFIB + OFE (n-n-h) *Total*: 744 +4(OFE) bytes |
| GEANT2 | 12 FNI + OFE (*Ref*) *Total*: 36 + OFE bytes | 76 nFIB + OFE (n-n-h) *Total*: 912 +4(OFE) bytes |
| Level 3 | 22 FNI + OFE (*Ref*) *Total*: 66 + OFE bytes | 184 nFIB + OFE (n-n-h) *Total*: 2208 +4(OFE)bytes |
| Sprint | 9 FNI + OFE (*Ref*) *Total*: 27 + OFE bytes | 60 nFIB + OFE (n-n-h) *Total*: 720 +4(OFE) bytes |

The third evaluation is an analysis of the FEP-S operation on a simulated online environment. The NotVia is not analyzed here because they are already known to work. Some changes have to be made in the original OSPF implementation of JavaSim to achieve the sub-second convergence [5]. With these changes, the OSPF can reduce its convergence time to approximately 200ms. During a failure occurrence, we have also changed the simulator to signal the OSPF or FEP-S only after 20ms to simulate hardware delay [5] (before this time, the packets are dropped). The objective is to evaluate the reduction of packet loss rate when the FEP-S is used compared to OSPF with sub-second convergence approach. We use GEANT2 topology representation, illustrated by Figure 3, which obeys the constraints of Section 2 (all routers remain connected in a single failure presence). GEANT2 was also tested with a reduced OSPF convergence period of 200 ms [5].  All links are projected in this evaluation to have traffic flows using at most 50% of their capacity. The link costs were set with a value inversely proportional to the link capacities (10Gbps). Among the traffic flows, the tests are focused on these six traffic flow pairs (origin, destination): (2,18), (18,2), (1,17), (17,1), (9,11) and (11,9). The shortest paths of these six traffic flows use routers 6 and 8. All these flows are generated with a constant bit rate of 1,6 Gbps with packets of 256 bytes to adjust the full duplex link 6-8 to use up to 4.8 Gbps (< 50% link capacity).

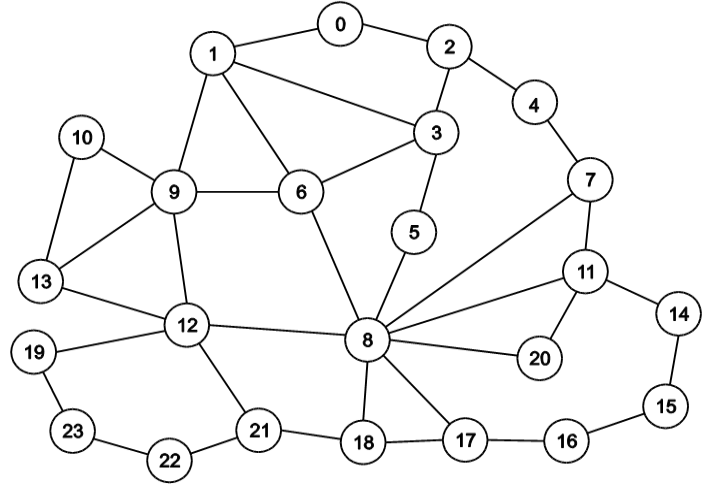

Figure 3. GEANT2 Network Topology

We design three independent failure scenarios: link 6-8, router 6 and router 8. These scenarios aim to analyze the failure of the central routers of this topology. First we made the tests with the modified OSPF and than using FEP-S combined with the modified OSPF, to aid it during the convergence. All the packets from the traffic flows analyzed, when deviated to other links, could be adjusted in the remaining bandwidth (50% free). This scenario enables an isolated packet loss rate analysis for the convergence period, which can not be achieved during the existence of high unplanned network traffic. However, during the *FEP_DifFor* deviation process, FEP-S is able to avoid higher network instability and congestion because it only continues forwarding the deviated marked packets (with *FEP_Mark*) if the queue length is shorter than a second (200ms). A similar approach to avoid congestion does not exist in any IPFRR approach, but it is a recommended action [7]. Moreover, this option can be enabled/disabled by the network administrator, because it is possible to have some priority traffic that should not be affected by a deviated traffic.

Table 5 presents the percentage of packet loss during the convergence period. The OSPF have a higher packet loss rate in all cases because of the convergence period and routing loops, even being it smaller than a second (200ms). The FEP-S helps the OSPF during its convergence time, because during this 200ms the *FEP_DifFor* is used to correctly forward the packets based on the *FEP_Mark/NI* until it reaches *RF*, which explains the lower packet loss rate. After 200ms, the routers have updated their OSPF routes and the *FEP_DifFor* is no longer needed. However, even so, the *FEP_DifFor* is programmed to maintain the forwarding process for a sufficient interval (Section 2.2.2) in order to guarantee the OSPF update in all affected routers.

Table 5. Packet loss rate percentage per traffic flow.

| Traffic Flow (Source, Destination) | OSPF (sub-second) | | | FEP-S | | |
|---|---|---|---|---|---|---|
| | Link 6-8 | Router 6 | Router 8 | Link 6-8 | Router 6 | Router 8 |
| (2,18) | 23,3% | 24,5% | 25% | 1,8% | 2,8% | 2,7% |
| (18,2) | 23,1% | 24,2% | 25% | 1,3% | 2,6% | 2,6% |
| (1,17) | 23,5% | 24,7% | 24,7% | 1,6% | 2,6% | 2,6% |
| (17,1) | 23,1% | 24,7% | 25% | 1,3% | 2,6% | 2,6% |
| (9,11) | 23,1% | 24,3% | 25,3% | 2,1% | 2,6% | 2,7% |
| (11,9) | 23,2% | 24,2% | 25,3% | 1,3% | 2,6% | 2,6% |

The deviation process generated by *FEP_DifFor* is based on *FEP_Mark* until the packets reach *RF*. From this router, the packets can safely be forwarded to *DR* with *FEP_DifFor*, based on the normal OSPF forwarding. In addition, the FEP-S recovery paths from *SR* to *DR* are almost equal to the OSPF, if it had converged to the failure. This occurs because FEP-S always generates recovery paths to bypass *AR*, if *AR≠DR*, which allows a gradual adaptation of the routers (*SR* to *RF*) to the new generated OSPF routes. The FEP-S percentage of packet losses occurs due to the hardware delay in detecting and reporting the failure. After this time, the *FEP_DifFor* process can act in order to deviate the packets.

## 4. CONCLUSIONS

In high speed IP networks, interior gateway protocols, like OSPF, cannot obtain a new route to bypass a failure in time. These protocols' convergence period can vary from hundreds of milliseconds to tens of seconds. This entails high packet loss rates until the end of this period, and it becomes worse in case of transient failure. We proposed the FEP-S approach to generate fast recovery paths in order to help routing protocols bypass failures during this period. The FEP-S obtained shorter recovery paths, used less FIB memory to identify these paths and is able to be useful in order to aid OSPF during its convergence period, avoiding high packet loss rate. Each recovery path obtained is almost always the same as that the OSPF would generate in case the router adjacent to a failure reacted. This allows a gradual adaptation of the recovery path to the OSPF path during the convergence period. Further tests with FEP-S, including an evaluation in a real environment, will be further provided. The security aspects of FEP-S and a modified approach to deal with multiple independent failures are also intended to be researched in future works.

## 5. APPENDIX A

An algorithm was developed for FEP-S Calculation and is presented in Table 6. This algorithm is a process to obtain the FEP_Vector[*DR*] and the respective *FEP_Mark/NI*. The router executing this calculation is named *SR*. In a given topology configuration maintained by OSPF, the algorithm reuses the *SR* OSPF shortest paths (step 1). In the sequence, the algorithm obtains the alternative paths ($\varphi_{SR,DR,AR}$) to each *DR* affected by a network interface simulated with failure, which can be (*SR,AR*) or *AR* (steps 2, 3, 4, 5 and 6). In the sequence, the algorithm searches for *RF* using the three classification levels to select the *S_FEP*$_{SR,DR,AR}$ (step 7). Finally, the FEP_Vector[*DR*] and the respective *FEP_Mark/NI* from *S_FEP*$_{SR,DR,AR}$ are obtained reusing the *FEP_Mark*s as possible (Sections 2.1 and 2.2).

Table 6.  FEP-S Calculation Algorithm.

| Steps | Description |
|---|---|
| 1 | Obtain the shortest paths resulted from normal OSPF calculation at *SR*. |
| 2 | For each active adjacent network interface at *SR* do: |
| 3 | If a fail is detected in the selected active adjacent network interface, identify which *DR* from step 1 would have their shortest path damaged. So, for each identified *DR* do: |
| 4 | Is *DR* equal to *AR* ? |
| 5 | If Yes: Remove the adjacent link (*SR,AR*) from topology and run the *Incremental-SPF* in order to find the *alternative paths* for *DR=AR*. If (*SR,AR*) belongs to a SRLG, all the links belonging to this SRLG are also removed from the topology in order to run *Incremental-SPF*. |
| 6 | Otherwise: Remove *AR* from topology and run the *Incremental-SPF* in order to find the *alternative paths* for all *DR* (obtained from step 3) different from *AR*. All remaining *DR* from step 3 should reuse these obtained *alternative paths* in |

| | order to avoid an unnecessary repeated *Incremental-SPF* run. If (*AR,?*) link belongs to a SRLG, all the links belonging to this SRLG are also removed from topology in order to run *Incremental-SPF*. The "*?*" symbol refers to any adjacent router to *AR*. |
|---|---|
| 7 | Reuse the results from step 5 or 6 in order to find an *emergency path* to RF according to the three classification levels (ECMP, LFA or SIG). Select the FEP with better classification level for each *DR* and store in FEP_Vector[*DR*]. |
| 8 | Generate the *FEP_Mark/NI* for each FEP_Vector[*DR*] according to Section 2.2 to reuse *FEP_Mark/NI* whenever possible to reduce FIB resource usage. |

**Authors**

**Fernando Barreto** received Phd degree in Sciences from Federal University of Technology – Paraná (Brazil). Recently is a teacher at Federal University of Technology – Paraná, campus Apucarana. The interest fields are network computers, network security and wireless networks.

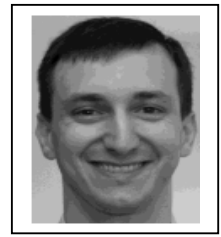

**Emilio C. G. Wille** received Phd degree at Tellecommunications and Electronic Engineering from Politecnico di Torino (Italia). Recently is an associated teacher at Electronic Department at Federal University of Technology – Paraná (Brazil). The interest fields are analysis and dimensioning of telecommunications network, queue theory, error control codes and optimization techniques.

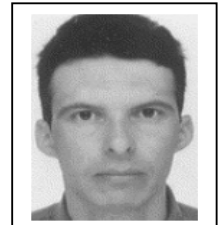

**Luiz Nacamura Junior** received Phd degree from Electric Engineering from Federal University of Santa Catarina (Brazil). Recently, is a teacher at Federal University of Technology – Paraná (Brazil). The interest fields are network computers, network management, distributed systems and Java language.

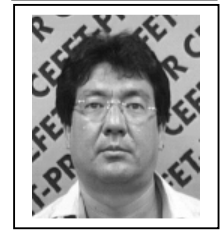